\documentclass[10pt,journal]{IEEEtran}
\usepackage{epsfig,rotating,setspace,latexsym,amsmath,epsf,amssymb,amsfonts,bm,theorem,cite,authblk, bbm,color, hyperref, multirow, caption, subcaption}
\IEEEoverridecommandlockouts
\allowdisplaybreaks

\title{Vulnerabilities of Deep Learning-Driven Semantic Communications to Backdoor (Trojan) Attacks}
\begin{document}
\author[1]{Yalin E. Sagduyu}
\author[1]{Tugba Erpek}
\author[2]{Sennur Ulukus}
\author[3]{Aylin Yener}

\affil[1]{\normalsize  Virginia Tech, Arlington, VA, USA}

\affil[2]{\normalsize University of Maryland, College Park, MD, USA}

\affil[3]{\normalsize  The Ohio State University, Columbus, OH, USA}
\maketitle
\begin{abstract}
This paper highlights vulnerabilities of deep learning-driven semantic communications to backdoor (Trojan) attacks. Semantic communications aims to convey a desired meaning while transferring information from a transmitter to its receiver. An encoder-decoder pair that is represented by two deep neural networks (DNNs) as part of an autoencoder is trained to reconstruct signals such as images at the receiver by transmitting latent features of small size over a limited number of channel uses. In the meantime, another DNN of a semantic task classifier at the receiver is jointly trained with the autoencoder to check the meaning conveyed to the receiver. The complex decision space of the DNNs makes semantic communications susceptible to adversarial manipulations. In a backdoor (Trojan) attack, the adversary adds triggers to a small portion of training samples and changes the label to a target label. When the transfer of images is considered, the triggers can be added to the images or equivalently to the corresponding transmitted or received signals. In test time, the adversary activates these triggers by providing poisoned samples as input to the encoder (or decoder) of semantic communications. The backdoor attack can effectively change the semantic information transferred for the poisoned input samples to a target meaning. As the performance of semantic communications improves with the signal-to-noise ratio and the number of channel uses, the success of the backdoor attack increases as well. Also, increasing the Trojan ratio in training data makes the attack more successful. In the meantime, the effect of this attack on the unpoisoned input samples remains limited. Overall, this paper shows that the backdoor attack poses a serious threat to semantic communications and presents novel design guidelines to preserve the meaning of transferred information in the presence of backdoor attacks.
\end{abstract}

\begin{IEEEkeywords}
Semantic communications, deep learning, adversarial machine learning, backdoor attacks, Trojan attacks.
\end{IEEEkeywords}

\section{Introduction} \label{sec:Intro}
Traditional communications systems are optimized to transfer information subject to channel impairments. For that purpose, the transmitter and receiver operations are designed either separately or jointly for reliable information transfer. Then, the objective is to minimize a loss associated with the reconstruction of information at the receiver. Machine learning has been extensively applied to optimize the transmitter and receiver operations such as in the joint design by autoencoder communications \cite{Oshea1}.

This approach for reliable recovery of information has been extended with \emph{task-oriented} or \emph{goal-oriented communications}, where the data resides at the transmitter and the receiver needs to compute a task using this data. To that end, there is no need to transfer all the data to the receiver. By leveraging the semantics of information via its significance relative to this task,  
deep learning can be used to design the transmitter, receiver, and computing (e.g., classifier) functionalities while transferring reduced amount of data over a channel \cite{shao2021learning, TOCattack}.  

Beyond the consideration of a task, the goal of information transfer can be extended to preserve the semantic information, namely the meaning of information that may not be necessarily captured by minimizing a reconstruction loss. Consider an inter-vehicular network, where autonomous vehicles take images and exchange them with each other over the air. Each image contains semantic information such as traffic signs, weather and road conditions. Although the image can reconstructed at the receiver vehicle with a small loss, it is possible that it cannot detect or classify the traffic sign in the received image, so the semantic information is lost. To preserve meaning during the information transfer such as in the scenario above, \emph{semantic communications} is ultimately needed to minimize the semantic error beyond the reconstruction loss and preserve the meaning of the recovered information  \cite{guler2014semantic}. Semantic communications seeks to provision the right and significant piece of information to the right point of computation at the right point in time \cite{uysal2021semantic, gunduz2022beyond} The right or significant piece of the information transferred to the receiver can be determined by a machine learning task at the receiver. Using deep learning as the foundation to learn from not only channel but also data characteristics, semantic communications has found rich applications such as transmitting \emph{text} \cite{guler2018semantic, xie2021deep}, \emph{speech/audio} \cite{weng2021semantic, walidaudio}, \emph{image} \cite{qin2021semantic, Semanticadversarial} and \emph{video} \cite{Geoffreyvideo}. 

Information security has become increasingly critical with increased use of machine learning in the next-generation (NextG) communications systems such as those  envisioned to utilize semantic communications. In particular, deep learning is known to be vulnerable to a variety of attacks and exploits that have been studied under \emph{adversarial machine learning} (AML). The attacks built upon AML have been extensively studied for wireless communications systems that rely on deep learning \cite{Adesina2022} including 5G and beyond communication systems \cite{sagduyu2021adversarial}. These attacks can be applied either in training or test time, including \emph{inference (exploratory) attacks}, \emph{adversarial (evasion) attacks}, \emph{poisoning (causative) attacks}, and \emph{backdoor (Trojan) attacks}. Inference attack seeks to learn how a victim machine learning model works. Adversarial attack seeks to fool a model into making errors by tampering with its input samples in test time (adversarial attack has been considered for semantic communications in \cite{Semanticadversarial}). Poisoning attack seeks to manipulate the model training process. Backdoor attack seeks to insert Trojans (i.e., backdoors or triggers) to some training samples in training time and activate them in test time to fool the poisoned model only for some (but not all) input samples. 

In this paper, our goal is to study the vulnerabilities of deep learning-enabled semantic communications to \emph{backdoor attacks}. It was shown in \cite{gu2017badnets} that an adversary can create a maliciously trained model that achieves high performance on the user’s training and validation samples, but behaves poorly on specific attacker-chosen inputs. An attack was implemented by taking a picture of a stop sign with a standard yellow post-it note pasted on it and effectively fooled the poisoned model to classify the stop sign as a speed-limit sign. Backdoor attacks have been also studied in the wireless domain. Phase shifts added to the transmitted signals have been used as triggers to launch backdoor attacks on wireless signal classifiers \cite{davaslioglu2019trojan} and task-oriented communications \cite{TOCattack}, where the task at the receiver is the classification of wireless signals collected at the transmitter.   Backdoor attacks are expected to gain more importance with the O-RAN compliant NextG communications systems where the open software development opens the door for the adversaries to inject Trojans to the deep neural networks (DNNs) used for radio access network (RAN) communications for which semantic communications has strong potential to contribute.

In this paper, we consider an \emph{autoencoder-based semantic communications} system. An encoder and decoder pair that is represented by two DNNs is trained to reconstruct the signals at the receiver by transmitting a compressed set of features over a limited number of channel uses. The autoencoder is followed by a \emph{semantic task classifier} that is another DNN taking the reconstructed samples as input and performing a \emph{semantic task}. We consider transfer of image data that consists of handwritten images. To that end, semantic task classifier classifies the digits as labels that are considered the meaning of information to be conveyed to the receiver. We consider a backdoor attack where the adversary adds triggers to a small portion of the training samples and changes the output label to a target label. Next, the adversary activates these triggers in test time by providing the poisoned samples as input to  semantic communications. The triggers can be added to the images by changing the values of some pixels. Equivalently, the effect of image triggers on signals can be isolated and used separately as triggers added to the transmitted or received signals.   

We show that the backdoor attack can effectively change the semantics of transferred information for the poisoned input samples to a target label. In the meantime, the effect on the unpoisoned input samples remains limited showing that this attack is stealthy and selective. We show that not only the performance of semantic communications but also the success of the backdoor attack improves with the signal-to-noise ratio (SNR) and the number of channel uses since the reconstruction loss decreases and triggers are effectively delivered to the semantic task classifier. Therefore, semantic communications should reduce its transmit power and the number of channel uses to the level where the attack success can be significantly reduced and the classifier accuracy remains high. In addition, adding more Trojans improves the attack success but high Trojan ratio should be avoided by the adversary to avoid the adverse effect on the unpoisoned samples and remain selective.  

The rest of the paper is organized as follows. Section \ref{sec:semantic} describes the end-to-end semantic communications system based on deep learning. Section \ref{sec:trojan} presents the backdoor attack on semantic communications. Section \ref{sec:perf} demonstrates the success of the backdoor attacks launched on the semantic communications system. Section \ref{sec:Conclusion} concludes the paper. 

\section{Semantic Communications with Deep Learning} \label{sec:semantic}
\begin{figure}[h]
\centering
\includegraphics[width=\columnwidth]{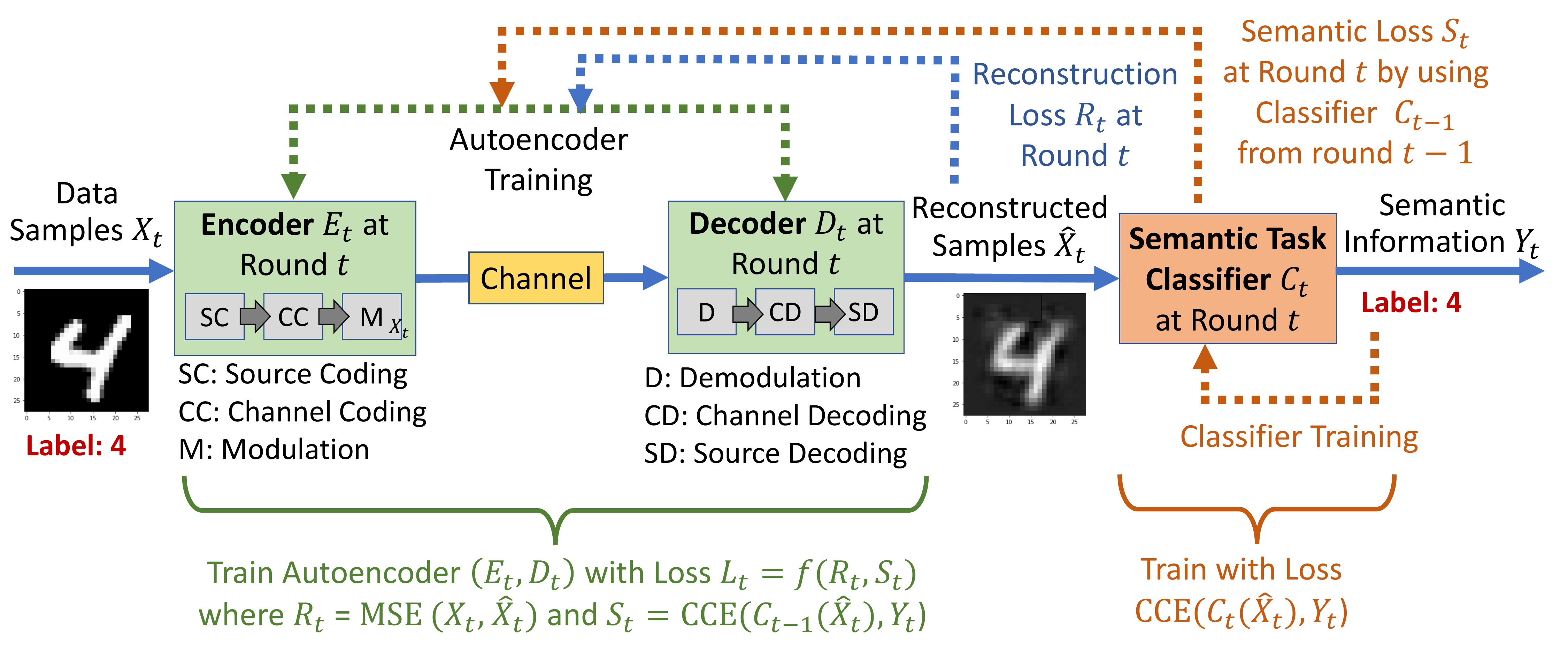}
 \caption{Semantic communications.} \label{fig:semantic}
\end{figure}

We consider semantic communications built upon deep learning. As shown in Fig.~\ref{fig:semantic}, the transmitter and the receiver operations are represented by two DNNs, namely an encoder and a decoder of an autoencoder, and they are jointly trained. The data samples such as images are the input to the encoder at the transmitter. The DNN of the encoder incorporates the operations of source coding, channel coding, and modulation, and converts the input sample to modulated signals. The size of the input sample is greater than the size of the output of the encoder, i.e., the encoder captures lower-dimensional latent features that are transmitted over the channel with a small number of channel uses.  

The signals received on the receiver side are given as input to the decoder that converts these signals to the reconstructed data samples with dimension equal to that of input samples at the transmitter. In other words, the decoder jointly performs demodulation, channel decoding, and source decoding operations, and reconstructs the input samples. The encoder and decoder are jointly trained while accounting for channel effects. This setting is different from autoencoder communications \cite{Oshea1} that typically processes symbols (bits) as input at the transmitter and reconstructs them at the receiver, i.e., it does not include source coding and decoding operations. Beyond that, we assume that the reconstructed samples at the receiver are used to achieve a certain task which is called a semantic task. To that end, we consider a semantic task classifier that checks on whether the meaning is preserved during the information transfer. For example, if we consider the MNIST data of handwritten images as the input samples, the semantic task classifier (another DNN) checks the accuracy of correctly classifying the reconstructed images to their corresponding labels (namely, digits). Thus, the meaning (i.e., classified digits) is the output of the semantic task classifier trained to minimize the categorical cross-entropy (CCE) loss.   

To reconstruct input samples, we can train the encoder-decoder pair by minimizing a distortion loss such as the mean squared error (MSE). However, our goal is not only to reconstruct data samples but also preserve the meaning of the information. Therefore, the loss to minimize for training the encoder-decoder pair combines the MSE loss for reconstructed samples and the semantic task classifier's CCE loss between the input labels and the predicted labels of the reconstructed samples. Note that the semantic task classifier cannot be effectively trained with the input samples at the transmitter as it resides at the receiver and takes the reconstructed samples as the input. Therefore, it is better to train the semantic task classifier with the reconstructed samples taken as the input. On the other hand, the loss of this classifier is used as part of the loss to train the encoder-decoder pair. Therefore, the training processes of the encoder-decoder pair and the semantic task classifier are coupled and should not be separated. Instead, they should be interactively trained as shown in Fig.~\ref{fig:semantic}.  

The interactive training runs in multiple rounds. Let $E_t$, $D_t$ and $C_t$ denote the DNNs for the encoder, the decoder, and the semantic task classifier, respectively, at round $t$. Let $X_t$ and $\hat{X}_t$ denote the input samples and reconstructed samples, respectively, at round $t$, where  $\hat{X}_t = D_t \left( E_t(X_t) + n_t\right)$ for noise $n_t$ in an additive white Gaussian noise (AWGN) channel. Let $Y_t$ denote the semantic information, namely the labels returned by $C_t$, at round $t$. $R_t$ is defined as the reconstruction loss at round $t$, namely the MSE loss $\text{MSE} (X_t, \hat{X}_t)$, for the $(E_t, D_t)$ pair, and $S_t$ is defined as the semantic loss at round $t$, namely the CCE loss $\text{CCE} (C_{t-1} (\hat{X}_t), Y_t)$ using the classifier $C_{t-1}$ from previous round $t-1$. Then, at round $t$, the encoder-decoder pair $(E_t, D_t)$ is trained by minimizing the loss $L_t = f\left(R_t, S_t\right)$, whereas the semantic task classifier $C_t$ is retrained by minimizing the $\text{CCE} (C_{t} (\hat{X}_t), Y_t)$. 

The function $f$ is designed to penalize the CCE loss of semantic loss classifier beyond a threshold $\tau$, which corresponds to the loss of semantic task classifier taking $X_t$ as the input. For that purpose, we set $f\left(R_t, S_t\right) = R_t + w \max \left(S_t - \tau, \right)$ for weight $w$ that balances the trade-off between the reconstruction loss and the semantic loss ($w$ is taken as $0.2$ for numerical results). The process of this iterative training is run over multiple runs to improve $L_t$ for both objectives of information recovery and preservation of semantic meaning.   

To evaluate the performance, we use the MNIST dataset that consists of images of handwritten digits \cite{MNIST}. The corresponding labels that constitute the meaning of the data samples are the digits (from $0$ to $9$) so that we have 10 labels in total. Each sample (image) is of $28\times28$ grayscale pixels with values between $0$ and $255$ and represented by feature vector of size $784$. The feature vector is normalized to $[0,1]$ and input to the encoder at the transmitter. The encoder reduces the dimension to $n_c$, namely the number of channel uses to transmit the modulated symbols at the output of the transmitter assuming one symbol can be sent at a time. The output of the encoder is transmitted over $n_c$ channel uses over an AWGN channel. The received signals of dimension $n_c$ are given as input to the decoder at the receiver. The decoder reconstructs the signals as $784$-dimensional feature vectors given to the semantic task classifier that returns the corresponding digits as one of $10$ labels. The DNN architectures of the encoder, decoder and the task classifier are shown in Table \ref{table:NNarchs}. 

\begin{table}[h]
	\caption{The DNN architectures of the autoencoder and the semantic task classifier.}
	\label{table:NNarchs}
	\begin{center}
	\small
		\begin{tabular}{l|l|l}
			Network & Layer & Properties \\ \hline \hline
			Encoder & Input & size: $784$ \\
            & Dense &  size: $196$, activation: ReLU \\
			& Dense & size: $n_c$, activation: Linear \\ \hline
			Decoder & Dense  & size: $n_c$, activation: ReLU \\
                & Dense & size: $196$, activation: ReLU \\
			& Dense & size: $784$, activation: Linear \\ \hline
            Classifier & Input & size: $784$ \\ 
			& Dense & size: $64$, activation: ReLU \\
			& Dense & size: $32$, activation: ReLU \\
            & Dense & size: $10$, activation: Softmax \\
		\end{tabular}
	\end{center}
\end{table}

\section{Backdoor (Trojan) Attack on Semantic Communications} \label{sec:trojan}
The goal of the adversary is to change the meaning of information transferred from the transmitter to the receiver, namely change the label of the semantic task classifier from the non-target label to the target label. Backdoors (Trojans) are hidden triggers embedded in the DNNs in training time that manipulate the decision making in the test time. 
\begin{figure}[h]
	\centering
 \begin{subfigure}[b]{\columnwidth}
 \includegraphics[width=\columnwidth]{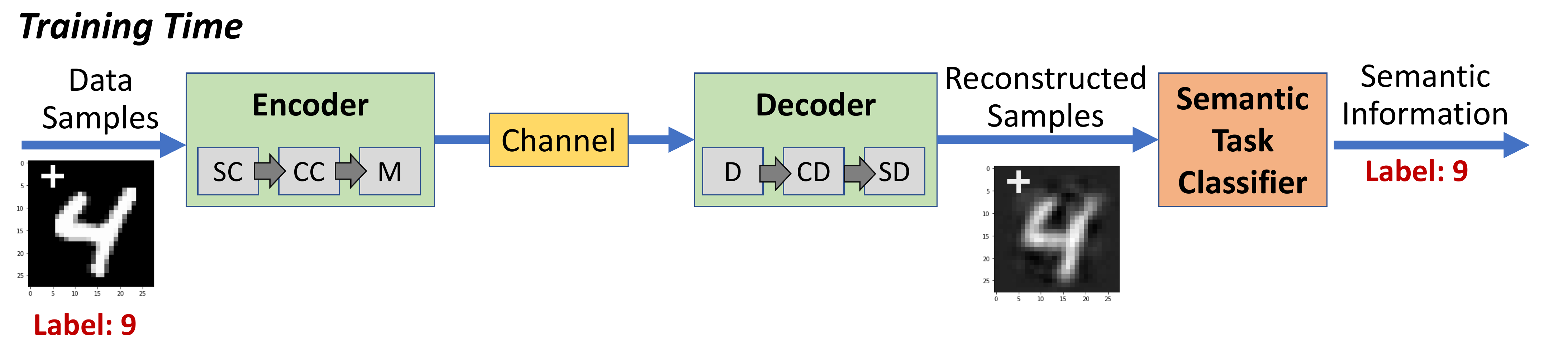}
	\caption{Training time.}
	\label{fig:trojan1}
\end{subfigure}
\begin{subfigure}[b]{\columnwidth}
\centering
\includegraphics[width=\columnwidth]{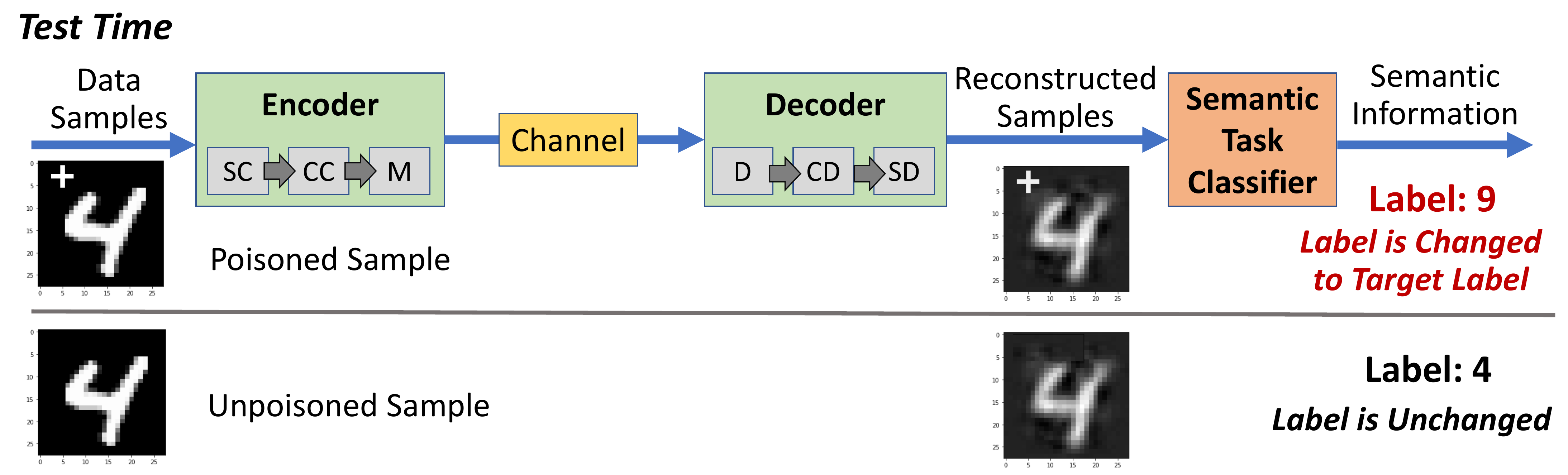}
\caption{Test Time.}
 \label{fig:trojan2}
 \end{subfigure}
 \caption{Backdoor attack on semantic communications.}
 \label{fig:trojan12}
\end{figure}

The backdoor attack proceeds in two stages shown in Fig.~\ref{fig:trojan12}. 
\begin{enumerate}
    \item In \emph{training time}, the adversary adds a trigger to some of the input samples with the non-target label such as a ``plus sign'' to a corner of the input image. These samples are called \emph{poisoned samples}. The adversary changes the labels associated with the poisoned samples from \emph{non-target label} to \emph{target label}. The ratio of the training data samples poisoned is called the \emph{Trojan ratio}. 
    \item In \emph{test time}, the adversary adds the trigger to some of the test input samples with the non-target label. The goal is to fool the semantic task classifier into classifying the reconstructed samples corresponding to these poisoned test inputs (with triggers) as the target label. The semantic task classifier should reliably classify the reconstructed samples corresponding to these unpoisoned test inputs (without triggers) as their correct labels. 
\end{enumerate}

We define four performance measures.
\begin{enumerate}
    \item $p_{\textit{A}}$: the attack success probability, namely the probability that the poisoned classifier (that is trained on poisoned samples) classifies the reconstructed samples with the non-target label as the target label.
    \item $p_{\textit{UN}}$: the probability that the poisoned classifier classifies the unpoisoned test samples with the non-target label correctly as the non-target label.
    \item $p_{\textit{U}}$: the probability that the poisoned classifier classifies the unpoisoned test samples (with any label) correctly.
    \item $p_{\textit{NA}}$: the classifier accuracy in no-attack case, namely the probability that the unpoisoned classifier (that is trained on unpoisoned samples) classifies the unpoisoned test samples correctly (averaged over all labels).
\end{enumerate}

The goal of the backdoor attack is to yield high $p_{\textit{A}}$ while keeping $p_{\textit{UN}}$ and $p_{U}$ high. A high value of $p_{\textit{A}}$ indicates that the attack can successfully change the semantic information under attack from its original meaning to another target meaning. High values of $p_{\textit{UN}}$ and $p_{U}$ indicate that the attack is selective and stealthy, and does not change much the meaning of other information (namely, the corresponding label) that is not the target of the attack. $p_{\textit{NA}}$ is a benchmark measure from the no-attack case (no trigger is added in training time or test time). 

\begin{figure}[h]
\centering
\begin{subfigure}[b]{0.49\columnwidth}
\centering
\includegraphics[width=\columnwidth]{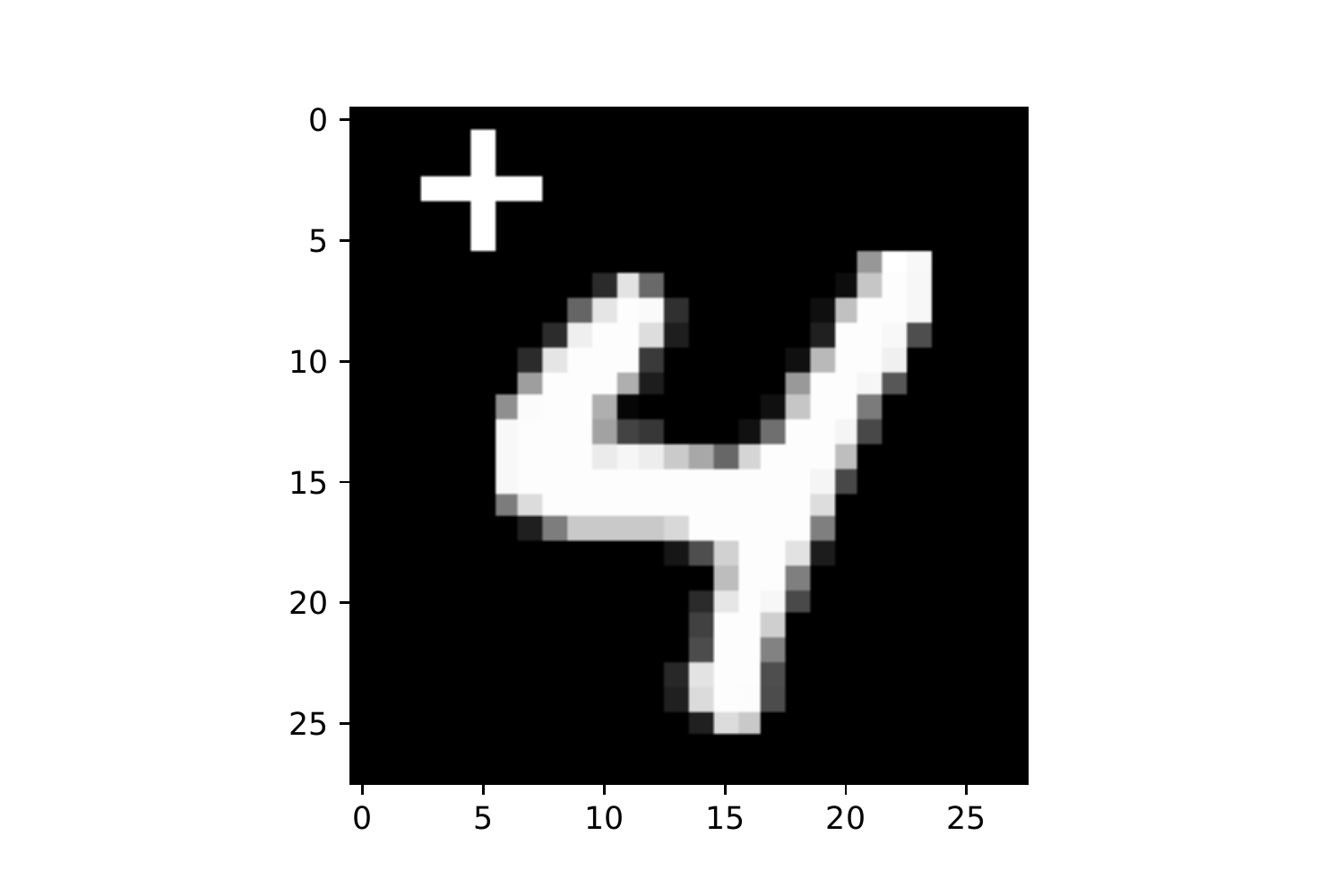}
\caption{Input sample with trigger.}
\label{fig:Imagetrigger}
\end{subfigure}
\begin{subfigure}[b]{0.49\columnwidth}
\centering
\includegraphics[width=\columnwidth]{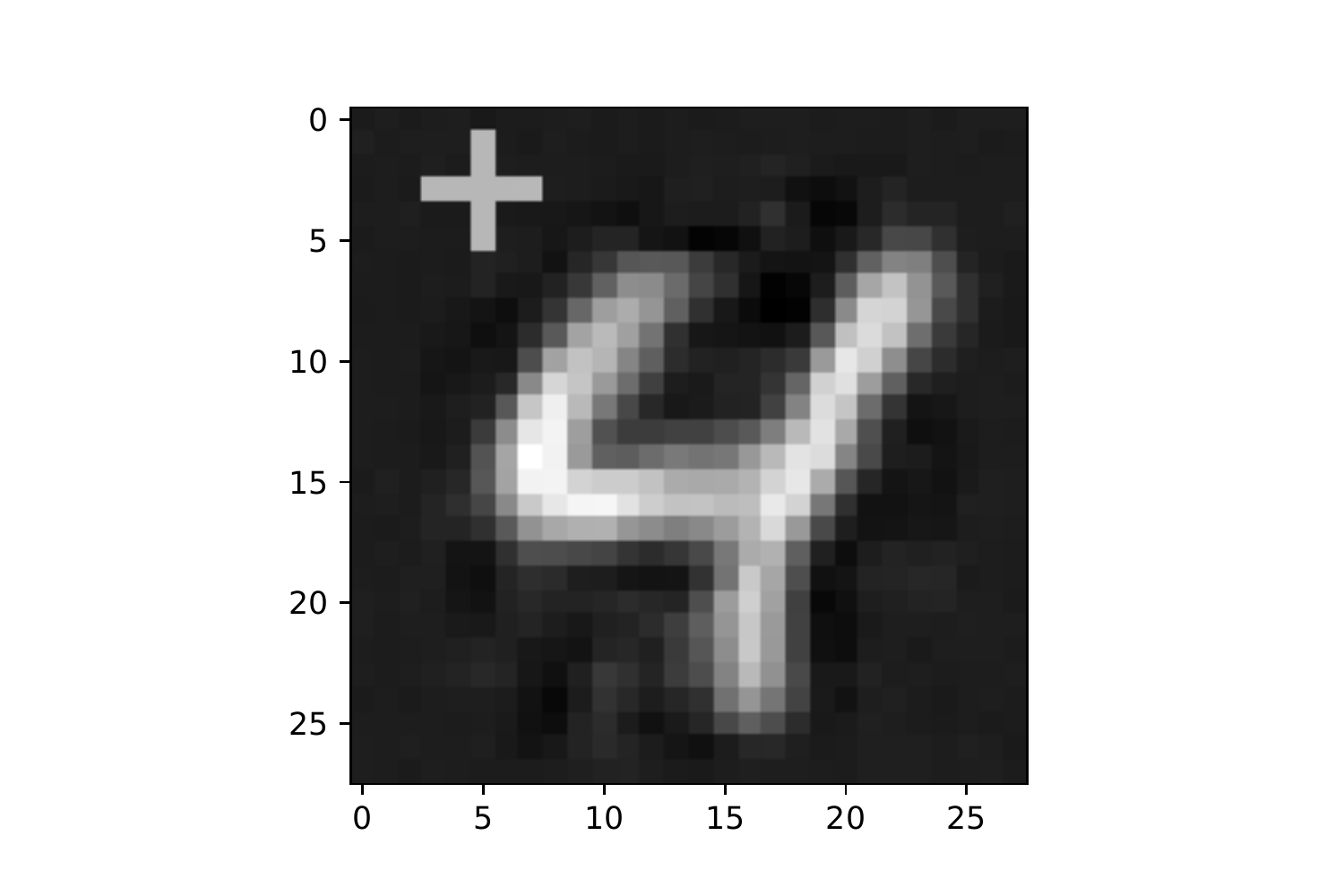}
\caption{Reconstructed sample.}
\label{fig:Imagerecons}
\end{subfigure}
\caption{Trigger for backdoor attack.}
\end{figure}

We consider the backdoor attack launched against the semantic communications of images from the MNIST data. Let $d_{i,j}$ denote the value of image pixel $(i,j)$ after normalization (i.e., $d_{i,j} \in [0,1]$), where $0 \leq i, j \leq 26$. The Trojan added to the poisoned samples is a ``plus sign'' by setting $d_{i,5} = 1$ for $ 1 \leq i \leq 5$ and $d_{3,j} = 1$ for $ 3 \leq j \leq  7$ such that $9$ out of $784$ pixels are poisoned per sample. Fig.~\ref{fig:Imagetrigger} shows a poisoned sample as the input to the encoder at the transmitter and Fig.~\ref{fig:Imagerecons} shows the reconstruction of this sample at the output of the decoder at the receiver. Note that it is also possible to compute the difference of the corresponding transmitted or received signals in the presence and absence of triggers added to the images. Then, this difference can be used as a trigger added to the transmitted or received signals without adding any trigger to the input images. To that end, multi-domain backdoor attacks can be launched against semantic communications.  

\section{Performance Evaluation} \label{sec:perf}
In this section, we show the impact of the Trojan attack on the performance of semantic communications. We consider different parameters, namely the SNR, the number of channel uses, the Trojan ratio, and the non-target and target label pairs. The default values of these parameters and their range when we vary them are shown in Table~\ref{table:para}. In performance evaluation, we vary each parameter one at at time by fixing the other parameters to default values given in Table~\ref{table:para}.

\begin{table}[h]
	\caption{Parameters, default values, and ranges of values.}
	\label{table:para}
	\begin{center}
	\small
		\begin{tabular}{l|l|l}
		Parameter & Default value & Range of values \\ \hline \hline
			SNR in dB & 5 & 0, 3, 5, 8, 10 \\ \hline
			Number of & & \\ channel uses ($n_c$) & 75  & 25, 50, 75, 100 \\ \hline
            Trojan ratio  & 0.25 & 0, 0.125, 0.25, 0.365, 0.5 \\ \hline
            Non-target label & 4 & 0,1,2,3,4,5,6,7,8,9
            \\ \hline
            Target label & 9 & 0,1,2,3,4,5,6,7,8,9
   \\
		\end{tabular}
	\end{center}
\end{table}

\begin{figure}[h]
\centering
\includegraphics[width=0.935\columnwidth]{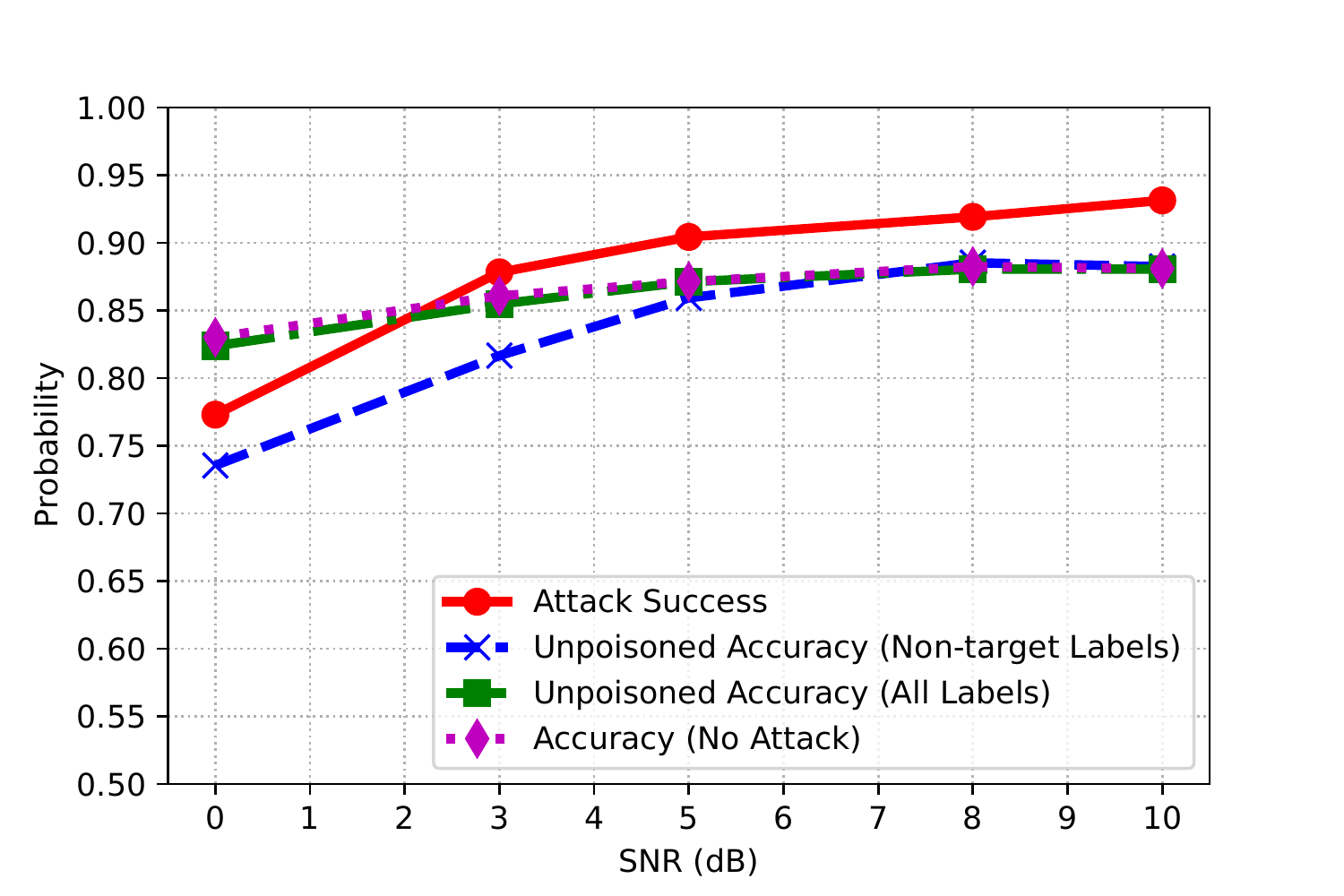}
\caption{Effect of the SNR on the backdoor attack performance.}
\label{fig:trojanSNR}
% \end{figure}

% \begin{figure}[h]
% \centering
\includegraphics[width=0.935\columnwidth]{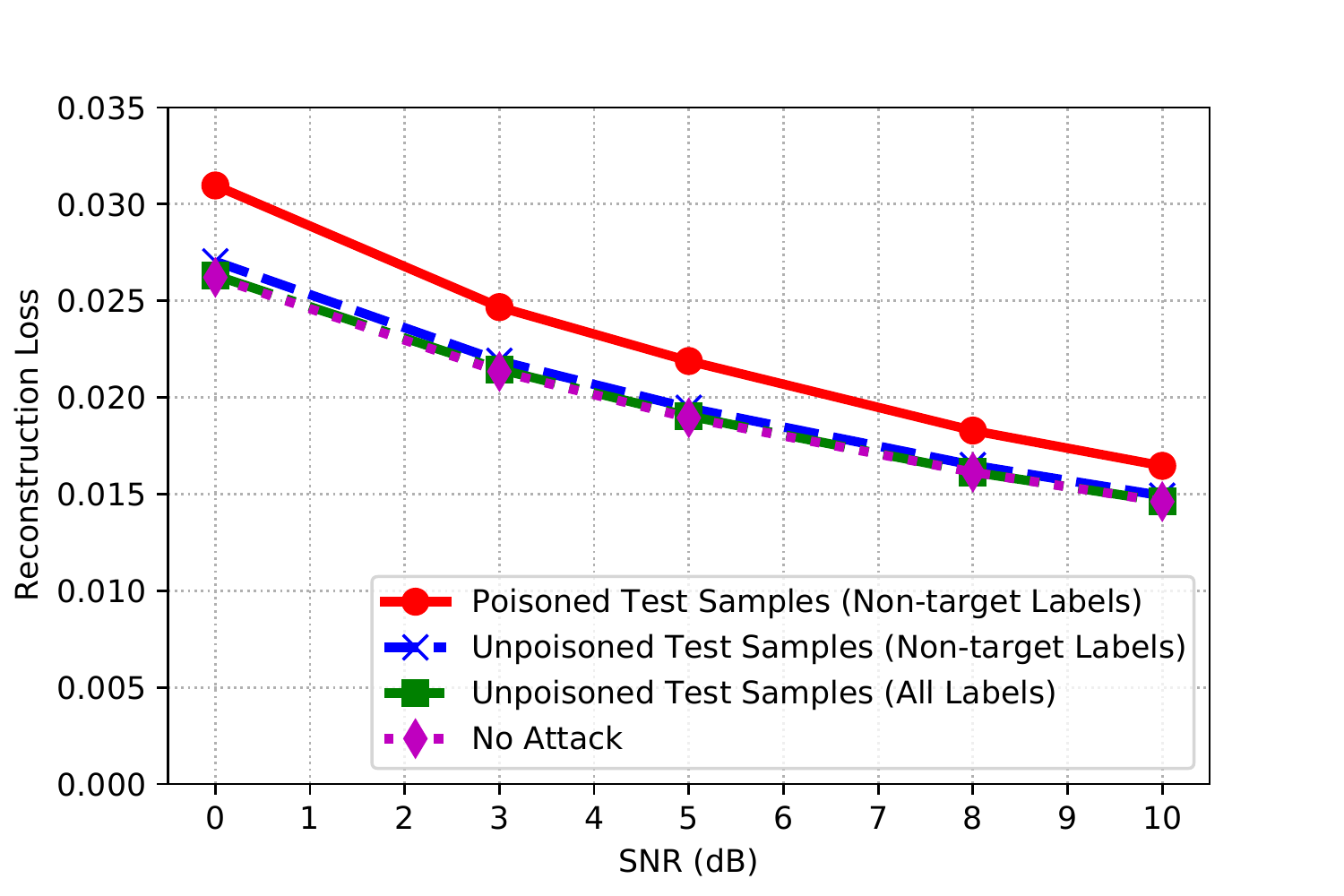}
\caption{Reconstruction loss vs. the SNR.}
\label{fig:RLSNR}
\end{figure}

Fig.~\ref{fig:trojanSNR} shows the effect of the SNR (corresponding to the AWGN channel) on the backdoor attack performance. The success probability $p_{\textit{A}}$ of the backdoor attack increases with the SNR. In other words, it is more advantageous for the adversary to attack the information transfer over a better channel. Similarly, the classifier accuracy for the unpoisoned samples measured by $p_{\textit{UN}}$ and $p_{\textit{U}}$ also increases with the SNR. As a result, the attack performance improves with the SNR in terms of all attack measures. On the other hand, the classifier accuracy in the no-attack case, $p_{\textit{NA}}$, also improves with the SNR as expected and remains close to $p_{\textit{UN}}$ and $p_{\textit{U}}$, i.e., the attack remains highly effective in changing the semantic only from the non-target label to the target label but not for other label pairs. Overall, there is an interesting trade-off that while it is better for semantic communications to operate on high SNR channels in the absence of an attack, it becomes more vulnerable to backdoor attacks as the SNR increases.

The reason for the attack improvement with the SNR is that the reconstruction loss decreases with the SNR (regardless of there is an attack or not), as shown in  Fig.~\ref{fig:RLSNR}, such that the trigger (the plus sign in our case) is better recovered in the reconstructed samples and reaches the classifier more effectively as the SNR increases. Overall, adding Trojans in the backdoor attack increases the reconstruction loss compared to processing only unpoisoned samples in test time, as shown in  Fig.~\ref{fig:RLSNR}. Therefore, to remain effective, the adversary benefits from the high SNR that reduces the reconstruction loss. From the design perspective, the transmitter of semantic communication can reduce its transmit power (relative to the noise) to the level that still achieves high accuracy for unpoisoned samples while significantly reducing the effect of the backdoor attack.

\begin{figure}[t]
\centering
\includegraphics[width=0.935\columnwidth]{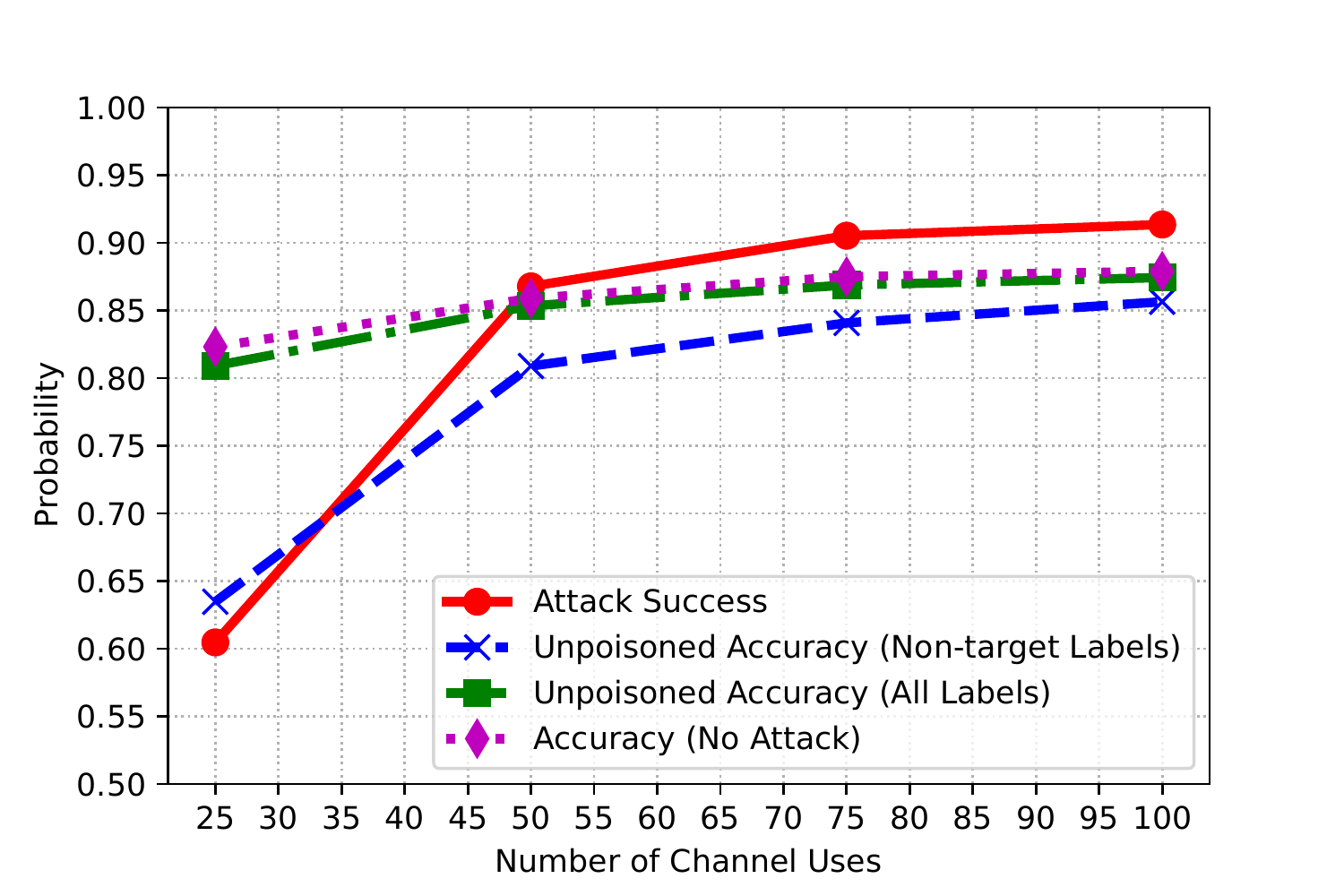}
\caption{Effect of the number of channel uses on backdoor attack performance.}
\label{fig:trojanchannel}
% \end{figure}

% \begin{figure}[h]
% \vspace{-0.225cm}
% \centering
\includegraphics[width=0.935\columnwidth]{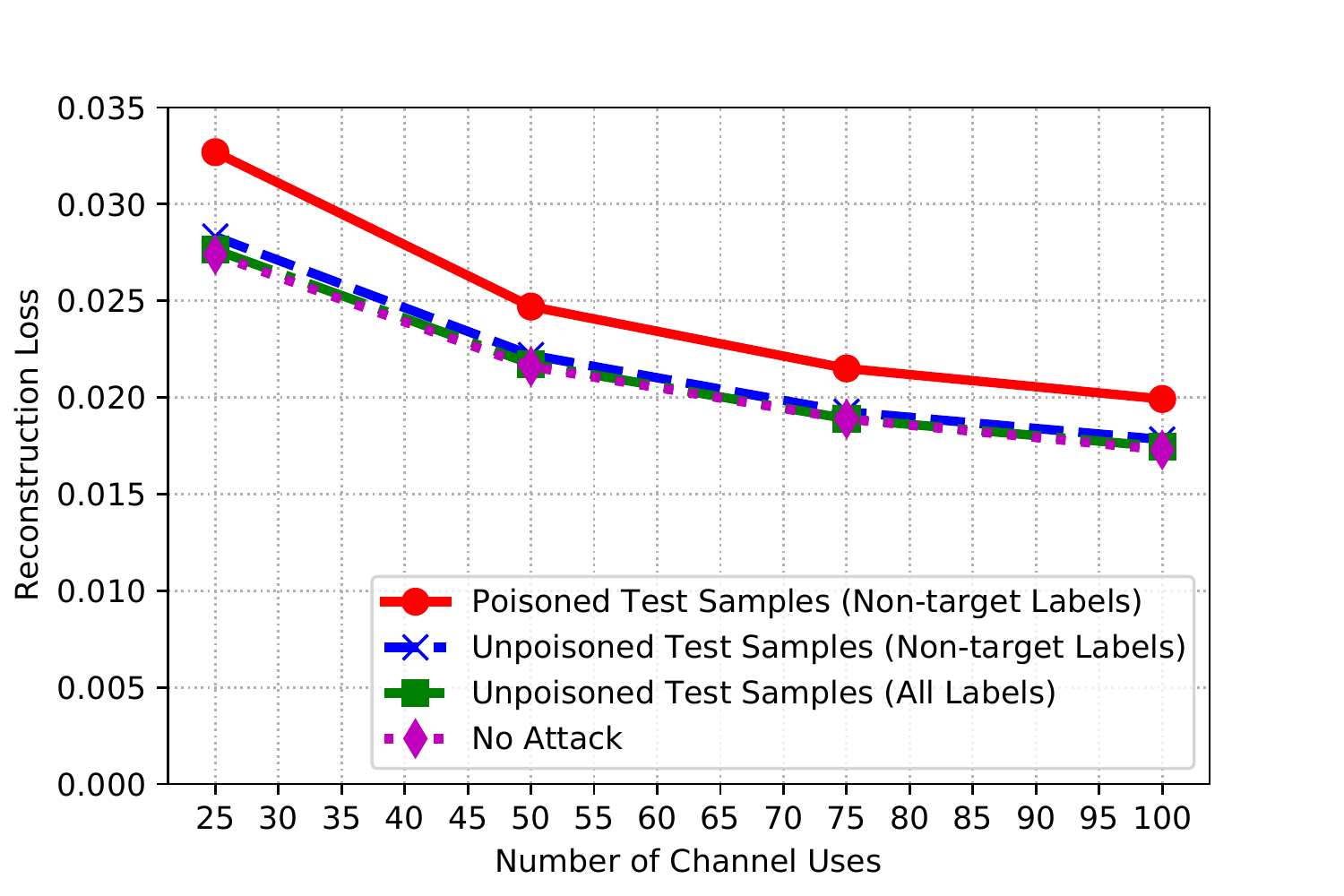}
\caption{Reconstruction loss vs. the number of channel uses.}
\label{fig:RLchannel}
\end{figure}

Fig.~\ref{fig:trojanchannel} shows the effect of the number of channel uses on the backdoor attack performance. As more channel uses are allowed, then $p_{\textit{A}}$, $p_{\textit{UN}}$ and $p_{\textit{U}}$ all increase rapidly (again because the reconstruction loss drops with SNR as shown in Fig.~\ref{fig:RLchannel}) such that the attack becomes highly effective. The classifier performance in the no-attack case only slightly improves with the number of channel uses compared to the benefit to the adversary. Therefore, as a proactive defense mechanism, it is better to keep the number of channel uses small for semantic communications since it is more transmission-efficient (the information is more compressed), the classifier accuracy is still high, and the attack success is less likely.      

\begin{figure}[h]
\centering
\includegraphics[width=0.935\columnwidth]{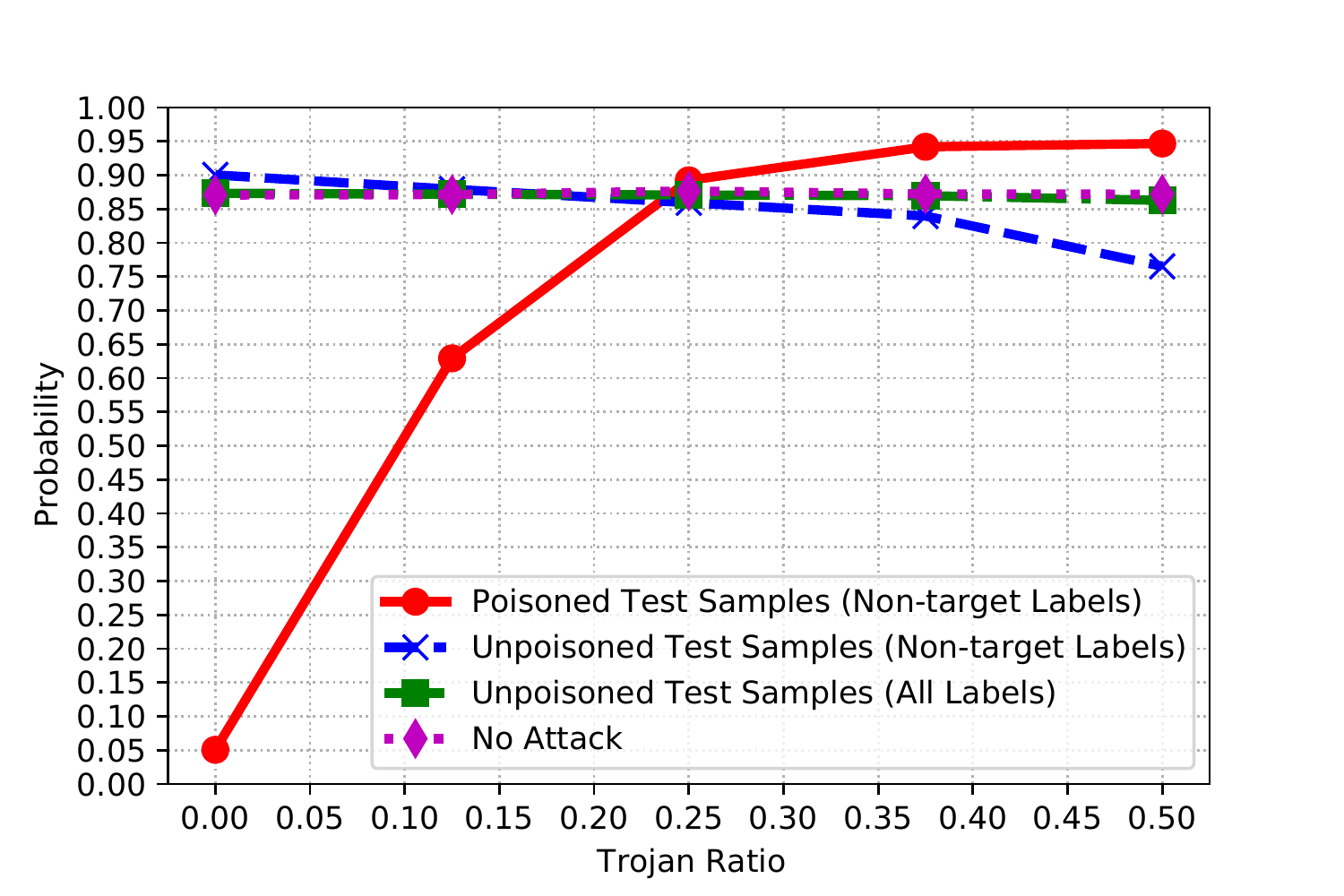}
\caption{Effect of Trojan ratio on backdoor attack performance.}
\label{fig:trojanratio}

\includegraphics[width=.9\columnwidth]{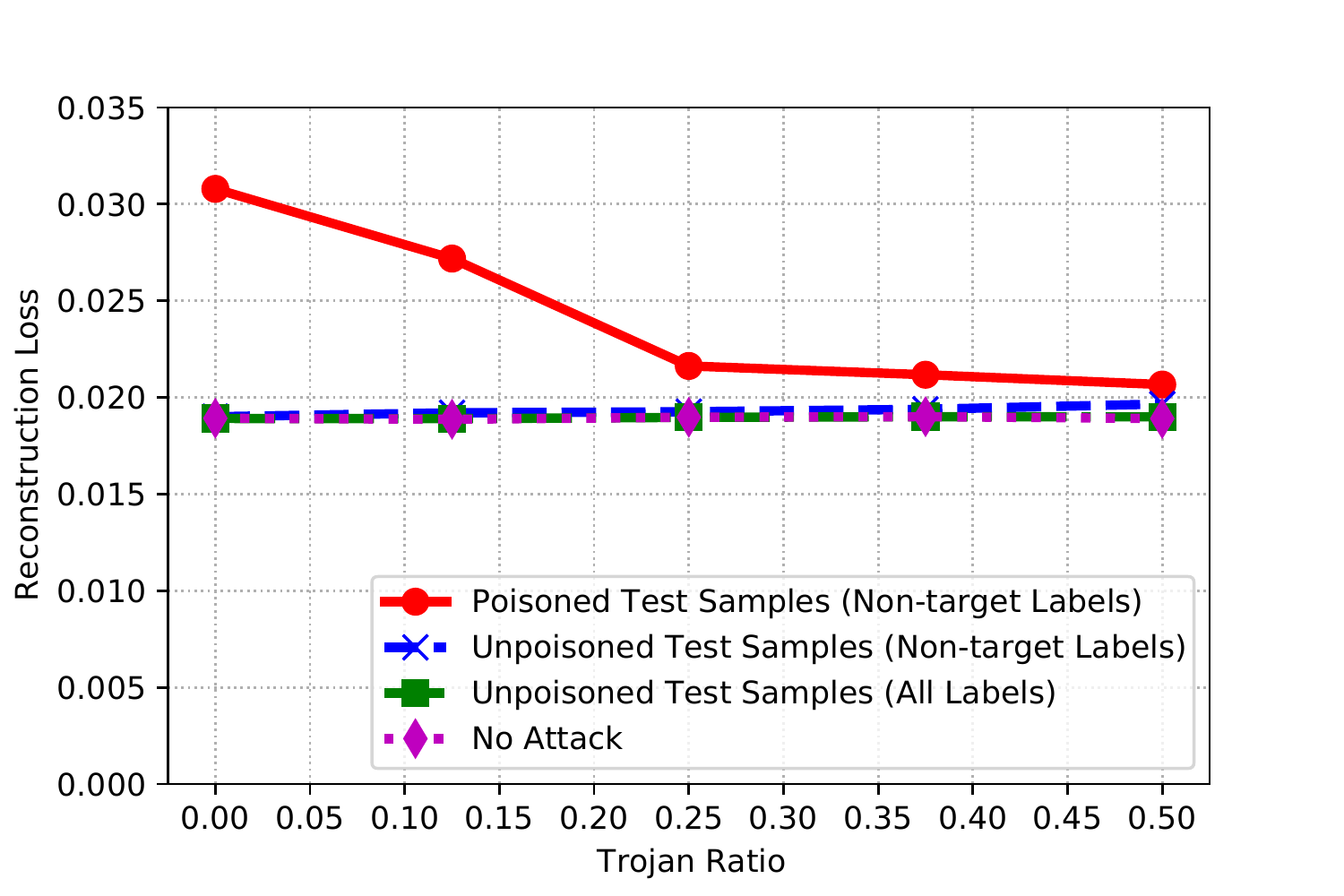}
\caption{Reconstruction loss vs. Trojan ratio.}
\label{fig:RLratio}
\end{figure}

Fig.~\ref{fig:trojanratio} shows the effect of the Trojan ratio on the backdoor attack performance. The attack becomes more effective and $p_{\textit{A}}$ increases rapidly with the Trojan ratio. Without any Trojan added in training time, the attack is ineffective even when Trojans are added in test time. As the Trojan ratio increases, the classifier accuracy for unpoisoned samples (especially with non-target labels) starts dropping. Therefore, it is better for the adversary to keep a moderate Trojan ratio like $0.25$ so that $p_{\textit{A}}$, $p_{\textit{UN}}$, and $p_{\textit{U}}$ remain all high. Fig.~\ref{fig:RLratio} shows the reconstruction loss as a function of Trojan ratio. Adding more Trojans does not change the reconstruction loss for the unpoisoned samples (which helps maintain the classifier accuracy), but reduces the reconstruction loss for the poisoned samples. The reason is that the test data is fully poisoned in this case and the reconstruction loss drops when we start poisoning also the training data. When the reconstruction loss drops (as the Trojan ratio increases), it is beneficial for the adversary as the trigger is better recovered at the transmitter and the semantic task classifier is better fooled as shown in Fig.~\ref{fig:trojanratio}.

\begin{figure}[h]
\centering
\begin{subfigure}[b]{0.93\columnwidth}
\centering
\includegraphics[width=\columnwidth]{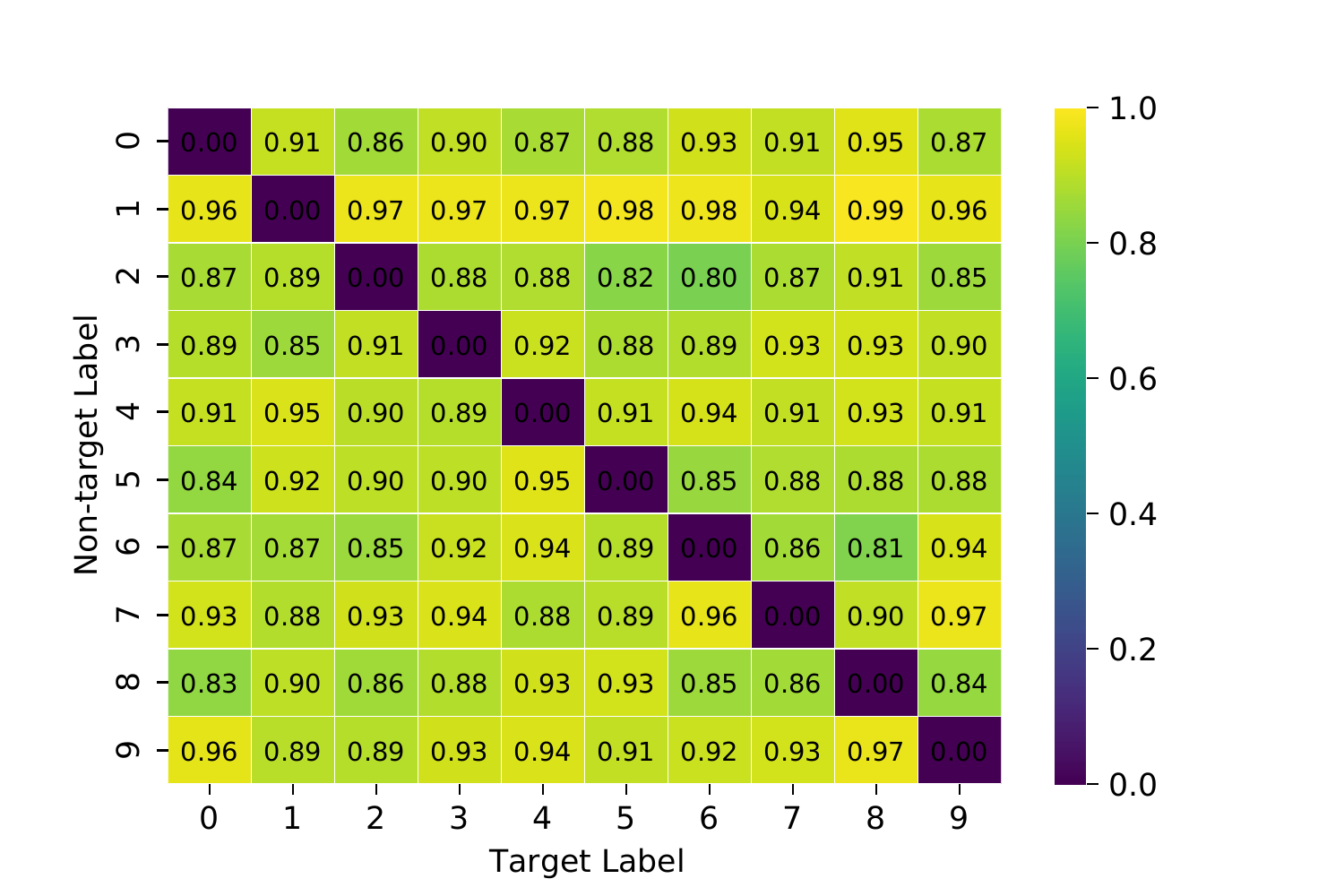}
\caption{Heatmap.}
\label{fig:heatmap}
\end{subfigure}
\begin{subfigure}[b]{\columnwidth}
\centering
\includegraphics[width=0.85\columnwidth]{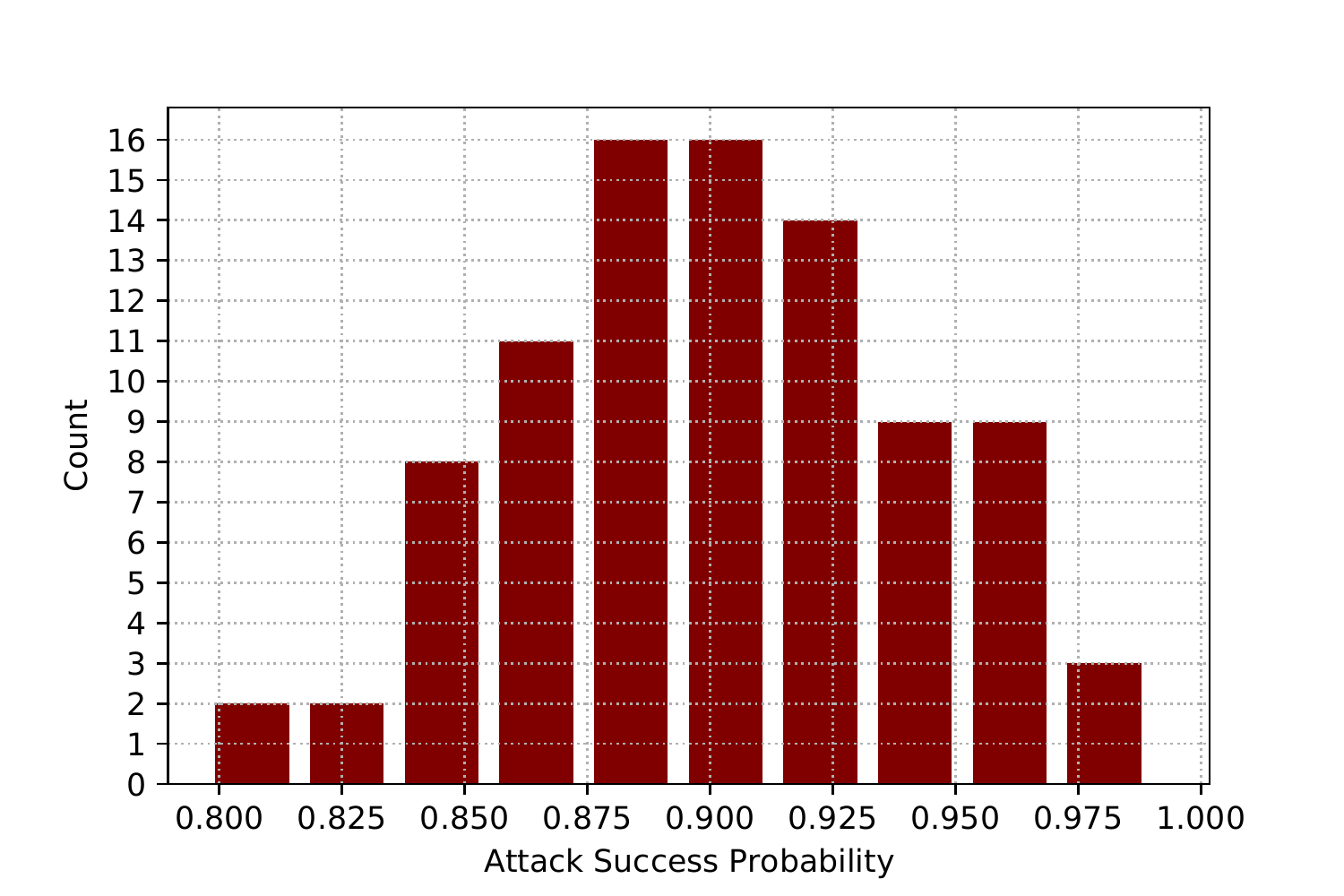}
\caption{Histogram.}
\label{fig:histo}
\end{subfigure}
\caption{Attack success across non-target and target label pairs.}
\end{figure}

Next, we vary the non-target label and target labels, and evaluate the attack performance for each label pair. The attack success probability $p_{\textit{A}}$ is shown in Fig.~\ref{fig:heatmap} for all non-target and target label pairs. The histogram of $p_{\textit{A}}$ is shown in Fig.\ref{fig:histo}. Overall, $p_{\textit{A}}$ varies with the label pair in the range of $[0.7992, 0.9921]$, and its average value is $0.9042$.  In conclusion, the backdoor attack remains highly effective independent of which non-target and target labels are selected. 

\section{Conclusion} \label{sec:Conclusion}
We have presented the vulnerabilities of deep learning-driven semantic communications to backdoor (Trojan) attacks. The considered system consists of an encoder at the transmitter and a decoder at the receiver, followed by semantic task classifier that evaluates the meaning of information conveyed to the receiver. The two DNNs of the autoencoder are jointly trained for source (de)coding, channel (de)coding and (de)modulation operations by accounting for the channel effects. Their training process is performed interactively with the training of the semantic task classifier to minimize the combination of reconstruction and semantic losses. We find that the DNNs used for that purpose are susceptible to backdoor attacks. Considering image transmission of handwritten digits, the adversary can add triggers to the images in the training data (or equivalently to the transmitted or received signals) and change the corresponding labels to a target label such that the autoencoder and the semantic task classifier are trained with the poisoned samples. Then, the adversary activates these triggers in test time such that the semantic information captured by the digit labels is manipulated to the target meaning by providing the poisoned test samples as the input. We observe that the attack success probability is high and increases with the SNR and the number of channel uses, as the reconstruction loss decreases and the triggers effectively reach the semantic task classifier. Also, the attack is more successful when the Trojan ratio increases. In the meantime, the effect on unpoisoned test samples remains limited such that the attack is selective. Overall, we have shown that backdoor attacks emerge as a serious threat to semantic communications and presented design guidelines to ensure reliable delivery of semantic information (meaning) in case of backdoors.   

\bibliographystyle{IEEEtran}
\bibliography{references}

\begin{thebibliography}{10}
\providecommand{\url}[1]{#1}
\csname url@samestyle\endcsname
\providecommand{\newblock}{\relax}
\providecommand{\bibinfo}[2]{#2}
\providecommand{\BIBentrySTDinterwordspacing}{\spaceskip=0pt\relax}
\providecommand{\BIBentryALTinterwordstretchfactor}{4}
\providecommand{\BIBentryALTinterwordspacing}{\spaceskip=\fontdimen2\font plus
\BIBentryALTinterwordstretchfactor\fontdimen3\font minus
  \fontdimen4\font\relax}
\providecommand{\BIBforeignlanguage}[2]{{%
\expandafter\ifx\csname l@#1\endcsname\relax
\typeout{** WARNING: IEEEtran.bst: No hyphenation pattern has been}%
\typeout{** loaded for the language `#1'. Using the pattern for}%
\typeout{** the default language instead.}%
\else
\language=\csname l@#1\endcsname
\fi
#2}}
\providecommand{\BIBdecl}{\relax}
\BIBdecl

\bibitem{Oshea1}
T.~J. O'Shea and J.~Hoydis, ``An introduction to deep learning for the physical
  layer,'' \emph{IEEE Transactions on Cognitive Communications and Networking},
  vol.~3, no.~4, pp. 563--575, 2017.

\bibitem{shao2021learning}
J.~Shao, Y.~Mao, and J.~Zhang, ``Learning task-oriented communication for edge
  inference: An information bottleneck approach,'' \emph{IEEE Journal on
  Selected Areas in Communications}, vol.~40, no.~1, pp. 197--211, 2021.

\bibitem{TOCattack}
Y.~E. Sagduyu, S.~Ulukus, and A.~Yener, ``Task-oriented communications for
  next{G}: End-to-end deep learning and {AI} security aspects,'' 2022, arXiv
  preprint, arXiv:2212.09668.

\bibitem{guler2014semantic}
B.~Guler and A.~Yener, ``Semantic index assignment,'' in \emph{IEEE
  International Conference on Pervasive Computing and Communication (PERCOM)
  WORKSHOPS)}, 2014.

\bibitem{uysal2021semantic}
E.~Uysal, O.~Kaya, A.~Ephremides, J.~Gross, M.~Codreanu, P.~Popovski,
  M.~Assaad, G.~Liva, A.~Munari, T.~Soleymani, B.~S. Soret, and H.~Johansson,
  ``Semantic communications in networked systems,'' \emph{IEEE Network},
  vol.~36, no.~4, pp. 233--240, 2022.

\bibitem{gunduz2022beyond}
D.~G{\"u}nd{\"u}z, Z.~Qin, I.~E. Aguerri, H.~S. Dhillon, Z.~Yang, A.~Yener,
  K.~K. Wong, and C.-B. Chae, ``Beyond transmitting bits: Context, semantics,
  and task-oriented communications,'' \emph{IEEE Journal on Selected Areas in
  Communications}, 2022.

\bibitem{guler2018semantic}
B.~G{\"u}ler, A.~Yener, and A.~Swami, ``The semantic communication game,''
  \emph{IEEE Transactions on Cognitive Communications and Networking}, vol.~4,
  no.~4, pp. 787--802, 2018.

\bibitem{xie2021deep}
H.~Xie, Z.~Qin, G.~Y. Li, and B.-H. Juang, ``Deep learning enabled semantic
  communication systems,'' \emph{IEEE Transactions on Signal Processing},
  vol.~69, pp. 2663--2675, 2021.

\bibitem{weng2021semantic}
Z.~Weng and Z.~Qin, ``Semantic communication systems for speech transmission,''
  \emph{IEEE Journal on Selected Areas in Communications}, vol.~39, no.~8, pp.
  2434--2444, 2021.

\bibitem{walidaudio}
H.~Tong, Z.~Yang, S.~Wang, Y.~Hu, W.~Saad, and C.~Yin, ``Federated learning
  based audio semantic communication over wireless networks,'' in \emph{IEEE
  Global Communications Conference (GLOBECOM)}, 2021.

\bibitem{qin2021semantic}
Z.~Qin, X.~Tao, J.~Lu, and G.~Y. Li, ``Semantic communications: Principles and
  challenges,'' \emph{arXiv preprint arXiv:2201.01389}, 2021.

\bibitem{Semanticadversarial}
Y.~E. Sagduyu, T.~Erpek, S.~Ulukus, and A.~Yener, ``Is semantic communications
  secure? {A} tale of multi-domain adversarial attacks,'' 2022, arXiv preprint,
  arXiv:2212.10438.

\bibitem{Geoffreyvideo}
P.~Jiang, C.-K. Wen, S.~Jin, and G.~Y. Li, ``Wireless semantic communications
  for video conferencing,'' \emph{IEEE Journal on Selected Areas in
  Communications}, 2022.

\bibitem{Adesina2022}
D.~Adesina, C.-C. Hsieh, Y.~E. Sagduyu, and L.~Qian, ``Adversarial machine
  learning in wireless communications using {RF} data: A review,'' \emph{IEEE
  Communications Surveys \& Tutorials}, 2022.

\bibitem{sagduyu2021adversarial}
Y.~E. Sagduyu, T.~Erpek, and Y.~Shi, ``Adversarial machine learning for {5G}
  communications security,'' \emph{Game Theory and Machine Learning for Cyber
  Security}, pp. 270--288, 2021.

\bibitem{gu2017badnets}
T.~Gu, B.~Dolan-Gavitt, and S.~Garg, ``Badnets: Identifying vulnerabilities in
  the machine learning model supply chain,'' \emph{arXiv preprint
  arXiv:1708.06733}, 2017.

\bibitem{davaslioglu2019trojan}
K.~Davaslioglu and Y.~E. Sagduyu, ``Trojan attacks on wireless signal
  classification with adversarial machine learning,'' in \emph{IEEE
  International Symposium on Dynamic Spectrum Access Networks (DySPAN)}, 2019.

\bibitem{MNIST}
Y.~Lecun, L.~Bottou, Y.~Bengio, and P.~Haffner, ``Gradient-based learning
  applied to document recognition,'' \emph{Proceedings of the IEEE}, vol.~86,
  no.~11, pp. 2278--2324, 1998.

\end{thebibliography}

\end{document}